\documentclass{PoS}

\title{The $\pi^0,\eta, \eta^{'} -> \gamma \gamma^*$ Decay Rates and Radii}

\ShortTitle{The $\pi^0,\eta, \eta^{'} -> \gamma \gamma^*$ Decay Rates and Radii}

\author{\speaker{A.M.  Bernstein}

        Physics Dept. and Lab. for Nuclear Science\\MIT, Cambridge Mass, USA.\\
        E-mail: \email{bernstein@mit.edu}}

\abstract{ The low $Q^2$  slopes of the transition form factors provide a unique method to measure the sizes of the neutral pseudo-scalar mesons, since they do not have electromagnetic form factors. From the slope one obtains the  "axial transition RMS radius" $ R_{PS,A} =  \sqrt{<r^2>}$ for each PS meson. The present status of theory and experiment for these quantities  are presented. A comparison of the $ R_{PS,A}$ is presented along with the electromagnetic and scalar radii of the $\pi^{\pm}$ mesons and  the proton. We observe the striking similarity  of the values of axial transition radii of all of the pseudoscalar mesons to each other and to the charge radius of the $\pi^{\pm}$. \\

~~~In the $Q^2$ = 0 limit the transition form factor is a measure of the pseudo-scalar meson radiative width (lifetime) and is a possible fourth (unexploited) method to perform such a measurement. The $\pi^{0} \rightarrow \gamma \gamma$ decay rate is a test of QCD at the confinement scale. There is a firm QCD prediction with a theoretical uncertainty of $\simeq $ 1 \% which calls for an experimental test at the same level of accuracy. 
There are three methods that have been utilized to perform this measurement and the  present status of the experimental tests are outlined. The current accuracy  is significantly less than the theoretical uncertainty. The  efforts to improve this are briefly summarized. }

\FullConference{The 8th  International Workshop on Chiral Dynamics \\
                 29 June 2015 - 03 July 2015\\
                 Pisa, Italy}

\begin{document}


\section{ The Transition Radii of Pseudoscalar Mesons From  $ PS \rightarrow \gamma^{*} (Q^2)  \gamma $ Decays}

The three pseudoscalar(PS) Nambu-Goldstone Bosons, $\pi^{0},\eta, \eta^{'}$, are neutral and due to charge conjugation symmetry do not have electromagnetic form factors. This means that we cannot directly measure one of their most important properties,  their physical size or RMS radius, as we can for charged pions, nucleons, etc. The closest we an come to such a measurement is the transition form factors $PS \rightarrow \gamma^{*} (Q^2)  \gamma $ at low $Q^{2},  F(Q^2) = F_{PS}(0) ( 1- Q^2 <r^2>/ 6 +.....)$, where the radiative width  $\Gamma (PS \rightarrow \gamma \gamma)= \pi m_{PS}^3 \alpha^2 F(0)^2/4$ (discussed in Sec. 2). From the slope of $F(Q^2)/F(0)$ at $Q^2$ = 0 we can obtain the "transition axial radius" $ R_{PS,A} =  \sqrt{<r^2>}$ for each PS meson. In recent times not much attention has been payed to this fundamental quantity. This may be in part because our physical intuition is guided by a non-relativistic understanding of hadronic densities and the interpretation of the RMS radius in models is subject to uncertainties due to the relativistic shifts of the reference frame. However the definition given here is model independent; it is equivalent to the slope of the form factor at $Q^2$ = 0. Most important, it is the same definition as other measures of RMS radii which are also obtained in a similar fashion from the relevant form factors so that comparisons of these measures are meaningful. As will be shown there are significant differences between the charge, axial, and scalar RMS radii, and the underlying physics has not been sufficiently well explored. 

\begin{figure}
\begin{center}
\includegraphics [scale=0.65] 
{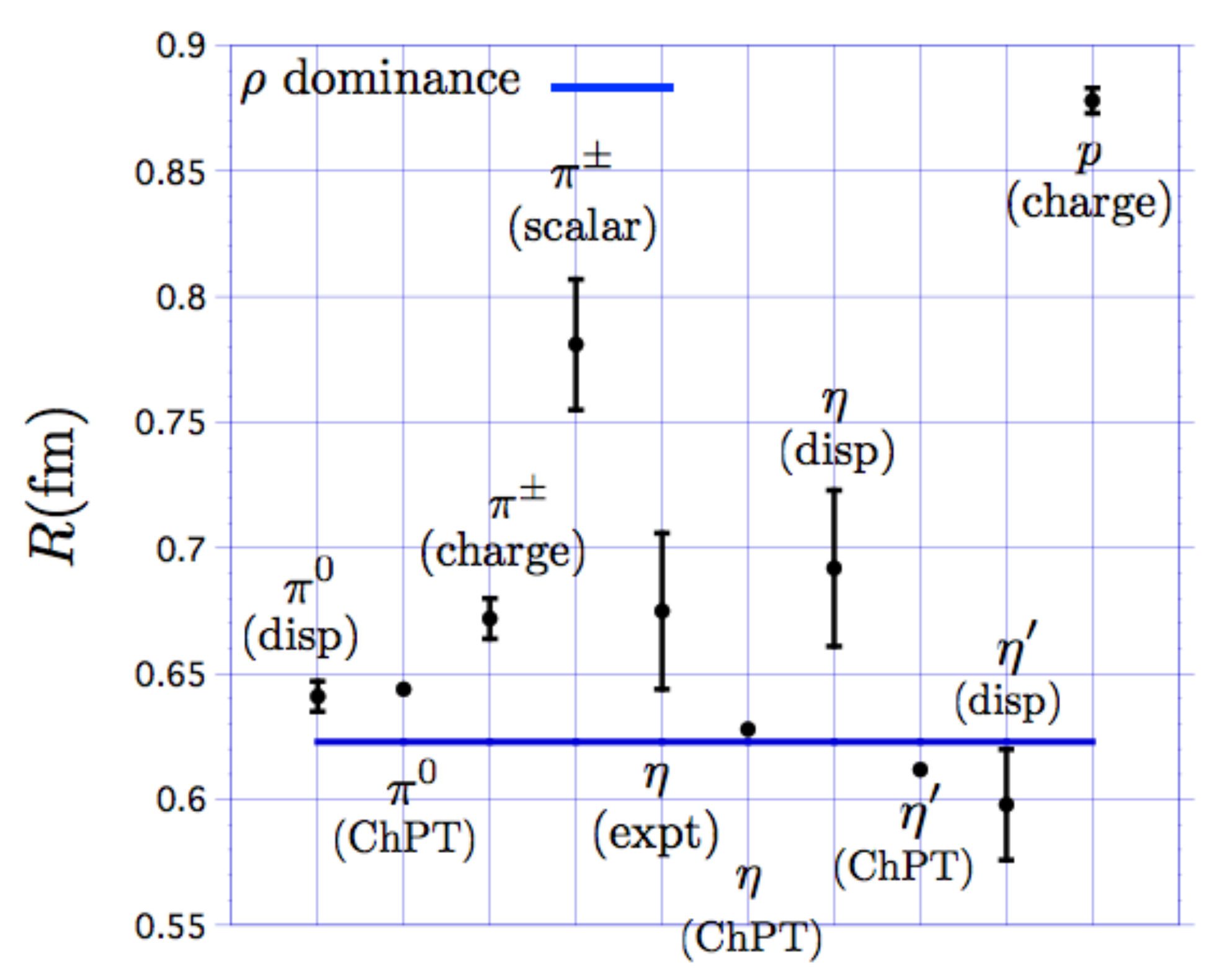}
\end{center}
\caption{RMS radii of the pseudoscalar mesons and the proton. From left to right: $R_{\pi^0, A}$ predicted from dispersion relations \cite{Martin}, $R_{\pi^0, A}$ predicted by ChPT \cite{R-ChPT}(no errors are quoted), the charge radius of the pion \cite{PDB}, the scalar radius of the pion \cite{Rpi-scalar}, $R_{\eta, A}$ measured by the $\eta \rightarrow \gamma^{*} (Q^2) \gamma$ reaction \cite {Reta-A2}, $R_{\eta, A}$ and $R_{\eta^{'}, A}$ predicted by ChPT \cite{R-ChPT}  and dispersion relations \cite{Reta-disp}, and the charge radius of the proton \cite{PDB}.Here the CODATA value is used, not the radius from muonic H which is $\simeq$ 0.04 fm smaller \cite{PDB}. The even larger scalar radius $\approx$1.3 fm obtained from a dispersion analysis of the form factor of the scalar "$\sigma$ term" \cite{R-pS}is not shown. The  horizontal line is the prediction of vector ($\rho$) dominance $ R = \sqrt{6}/m_{\rho}$ = 0.62 fm.  }
\label{fig:R}
\end{figure}

In Fig.\ref{fig:R} the RMS radii of the pseudoscalar mesons are presented and compared to the value predicted by vector (rho) resonance dominance $ R = \sqrt{6}/m_{\rho}$ = 0.62 fm which gives the right order of magnitude for most of the hadronic radii. It is the small deviations from this value that provides a clue to the hadron dynamics, which means that meaningful experiments must be accurate.  For the $\pi^{0} \rightarrow \gamma^{*}(Q^2) \gamma $ decay the kinematic range is limited so that the most accurate value comes from a dispersive theoretical treatment \cite{Martin}. It can be seen that the ChPT prediction \cite{R-ChPT} agrees with the dispersive calculation. The close agreement with the vector dominance is somewhat accidental and comes from the contributions of various terms \cite{R-ChPT}. It is interesting to note how similar the charged pion RMS radius \cite{PDB} is to the  $\pi^{0}$ axial transition radius, slightly higher by $(0.03 \pm 0.01)$ fm. On the other hand, the pion scalar radius, determined from a dispersive analysis of $\pi-\pi$ scattering \cite{Rpi-scalar}, is significantly higher ( 0.14$\pm$0.03 fm or 22$\pm$ 4\%; see the Appendix for a discussion). It is of interest that recent lattice calculations are in agreement with this value \cite{Lattice}.

The Mainz A2(real photon) group has recently measured the $\eta$ axial transition radius in the $\eta \rightarrow \gamma^{*}(Q^2) \gamma$ reaction \cite{Reta-A2}. Within the experimental error this result is in agreement with the predicted value based on dispersion relations \cite{Reta-disp} and the $\pi^{0}$ axial transition radius \cite{Martin}. This value is above the ChPT \cite{R-ChPT} and vector dominance prediction. For the $\eta^{'}$ there are no published experimental results in the low $Q^2$ region but the  Mainz A2 group has data which is being analyzed \cite{A2}. There are also plans to make such a measurement at JLab \cite{Mike}. There are predicted values based on dispersion relations \cite{Reta-disp} and ChPT \cite{R-ChPT}, both of which predict $R_{A,\eta^{'}} < R_{A,\eta}$.  A more recent calculation \cite{Kubis} predicts a reduction of 7\% in the $\eta$ form factor slope or 3.5\% in its transition radius \cite{Kubis}. This amounts to a reduction of 0.025 fm for the radius, approximately equal to the quoted error. In Fig. 1 the older prediction \cite{Reta-disp} is used because it is of interest to see the $\eta, \eta^{'}$ difference.  It will be interesting to see if experiments agree with the prediction that $R_{A,\eta^{'}} < R_{A,\eta}$ and to have a physical explanation  for this difference. 

A striking result shown in Fig.\ref{fig:R} is how similar the RMS radii are for the axial transition radii of all of the pseudoscalar mesons. In addition for the pion $R_{A,\pi^0} \simeq R_{charge, \pi^{\pm}}$.The experimental ratio $R_{A,\pi^0} /R_{charge, \pi^{\pm}} = 0.953 \pm 0.012$. Intuitively  this might not be expected since for the charge radius the contributions of the up and down quarks have opposite signs, whereas for the axial transition radius are proportional to the square of the quark charges. A back of the envelope estimate using simple quark model wave functions gives a ratio $R_{A,\pi^0} /R_{charge, \pi^{\pm}} \simeq 0.84$

It is of interest to compare the charge radius of the proton to the radii of the pseudoscalar mesons even though they have different quark substructures. From Fig. 2 it is seen that the charge radius of the proton is significantly larger than the radii of the pseudoscalar mesons and also the one predicted by vector (rho) dominance. There is a significant history of vector dominance calculations for the nucleon form factors.To achieve high quality agreement other mesons than the $\rho$ have to be included including the $\omega,\phi$ and some higher mass vector mesons \cite{VMD}. 

I hope that this discussion about the slope parameter of the $PS \rightarrow \gamma^{*}( Q^2) \gamma $ form factor stimulates new, accurate experiments and further calculations. In particular it is of interest to re-examine the ChPT calculations \cite{R-ChPT}, to extend the lattice calculations\cite{Lattice}, and perhaps most important, to physically interpret that differences between the charge, scalar, and axial transition RMS radii.  

\section{The $\pi^{0} \rightarrow \gamma \gamma$ Decay Rate.}
\label{sec:pi0 decay}
The  $\pi^{0}\rightarrow \gamma \gamma$ decay rate is dominated by
the QCD chiral anomaly \cite{anomaly}; this represents the explicit symmetry breaking by the electromagnetic field of the chiral symmetry associated with the third isospin component of the axial current \cite{anomaly}. The $\pi^{0}$ decay actually provides the  most sensitive test of this phenomenon of symmetry breaking due to the quantum fluctuations of the quark fields in the presence of a gauge field. 
In the limit of vanishing quark masses the anomaly leads to the $\pi^0\to \gamma\gamma$  decay amplitude \cite{anomaly}is predicted to be $\Gamma(\pi^{0} \rightarrow \gamma \gamma)= (\alpha/F_{\pi})^2 (m_{\pi^{0}}/4 \pi)^3 = 7.725 \pm 0.044 $eV with the  0.6\% uncertainty due to the experimental error in $F_{\pi}$, the pion decay constant \cite{PDB}.
This prediction, which is the dominant contribution to the $\pi^{0}$ decay rate, has no adjustable parameters. This  decay rate  is exact only in the chiral limit, {\sl i.e.}, when the $u$ and $d$~quark masses vanish. The chiral symmetry of QCD is explicitly broken by the finite quark masses. Since the masses of the up and down quarks are not equal this also leads to isospin breaking effects primarily due to $\pi^{0},\eta, \eta^{'}$ mixing.Three somewhat different chiral perturbation theory(ChPT)  calculations are in excellent agreement with each other and predict an increase of $\Gamma(\pi^{0} \rightarrow \gamma \gamma)$ of $4.5 \pm 1.0$ \% (see \cite{AB-review} for a discussion of the theory and also \cite{Rory-review} for references to the theory and experiments).This strong isospin breaking is larger than the typical values of $\simeq$ 1 to 2\%. Most important, this a firm QCD prediction that allows a test of this fundamental theory at the confinement scale, and which in turn sets an accuracy goal $\simeq$ 1\% for modern measurements. 

\begin{figure}
\begin{center}
\includegraphics [height=0.7\textwidth, width=0.7\textwidth] 
{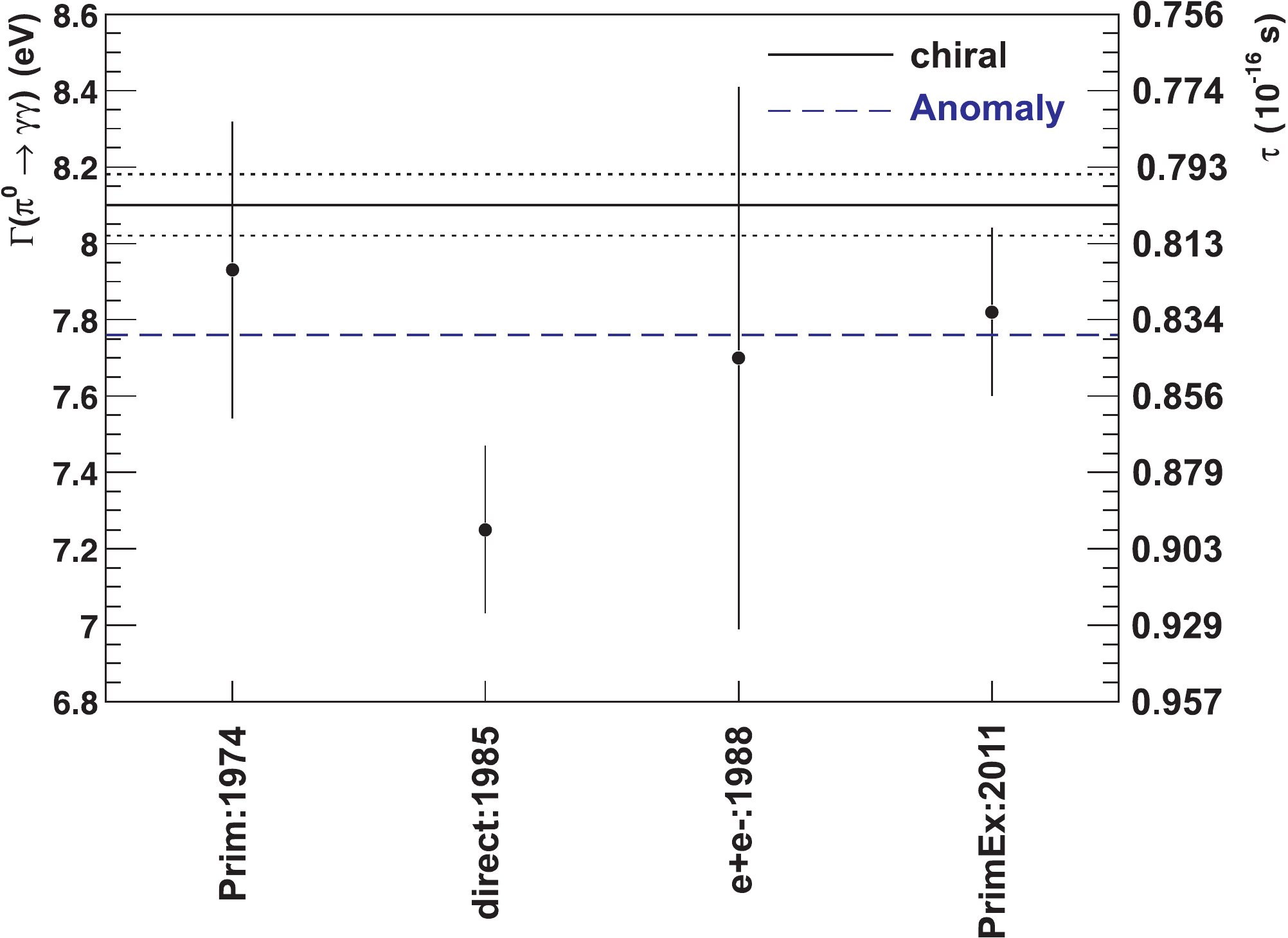}
\end{center}
\caption{$\Gamma(\pi^{0} \rightarrow \gamma \gamma)$) in eV(left scale) and lifetimes(right scale; note the suppressed zeros); see text for discussion. The lower dashed line is the result of the chiral anomaly and the  upper solid line with the dashed lines showing the theoretical 1\% error. For references to the theory and previous experiments see a recent review article \cite{AB-review} and for the experiments the particle data book \cite{PDB}.}
\label{fig:width}
\end{figure}

A comparison of theory and experiment is presented in Fig.\ref{fig:width}. There are three experimental methods: the direct measurement of the distance that high energy $\pi^{0}$s travel before decaying(1985), a Primakoff measurements performed at Cornell(1974) and recently at JLab(PrimEx1,2011), and  a two photon production cross section measurement in $e^{+} e^{-}$ collisions(1988). At the present time there are no significant deviations from the QCD prediction. However, it can be seen that  the experimental consistency and errors (some understated) are not sufficient to test this fundamental prediction. Fortunately improvements in the experiments are on the horizon. The JLab experiment is being repeated (PrimEx2,  the next talk by A.Gasparian in this session). The Compass collaboration at CERN is considering  an improved direct measurement of the recoil distance\cite{AB} and the DAPHNE group at Frascati is engaged in a more accurate measurement of the two photon $e^{+} e^{-}$ measurement \cite{Frascati}.

 Finally we note that there is one more experimental method that has not been explored, namely a measurement of the transition form factor in electron scattering (the virtual Primakoff effect) in the limit as $Q^2 \rightarrow $ 0. This would have a very different systematic errors than a measurement of the  Primakoff effect with real photons.  

\section{Appendix: The Scalar Radius}

Form factors depend on the structure of the particle being studied and the operator; \\ $ F_{O} (t) = < \psi(p^{'} | O | \psi(p) > $. where O is the specific operator, $\psi $ is the wave function of the particle, $p, p^{'}$ are the four momentum vectors of the initial and final states, and the four momentum transfer $ t = -Q^2 = (p^{'} - p )^2$. For the scalar form factor the operator $O_{S} \equiv \hat{m} (\bar{u} u +  \bar{d} d )$ where \\  $ \hat{m} = (m_{u} + m_{d})/2$ \cite{Rpi-scalar,R-pS}; this is a scalar in spin and isospin space. As was discussed in the introduction to Sec. 2 the RMS radius $R_{O}$ corresponding to each operator is obtained from the slope of $F(Q^2)$  at $Q^2=0$. If resonance dominance would work for scalar radii, for the pion we might anticipate that it is the $\sigma$ meson which has the right quantum numbers. This is a very unusual situation where the mass $\sim$ 440 MeV and the width is about half as large ( see \cite{PDB} and references therein). The pole is so far from the real axis that one cannot expect resonance dominance to pertain as it does for the $\rho$ meson which has a narrow width \cite{PDB}. Using the sigma mass one obtains $R_{\sigma, dominance } \sim 1.1$  fm. It is clearly naive to expect this estimate to be realistic in this case without a dispersion calculation which reflects the distance that the sigma is from the real axis \cite{Heiri}. This is clearly much larger than the value of $0.78 \pm 0.03$ fm \cite{Rpi-scalar}. Qualitatively however the relatively low mass of the sigma compared to the $\rho$ explains why this radius is larger, although as explained, it is not a quantitative  prediction. For the proton there is no corresponding low lying strong resonance with the same quantum numbers \cite{PDB} so on the basis of resonance dominance one expects an even larger scalar radius, which is what is found \cite{R-pS, Heiri}.

 Acknowledgements\\ I would like to thank B. Kubis and Z.Davoudi for their comments on the manuscript, and to  J. Friedrich, M. Hoferichter, B. Kubis, H. Leutwyler, M. Osrick, and M. Williams for valuable discussions. 


\begin{thebibliography}{99}
 \bibitem{Martin} M. Hoferichter et al., Eur. Phys. J. C74 (2014) 3180.
 \bibitem{R-ChPT} L. Ametller, J. Bijnens, A. Bramon, F. Cornet, Phys.Rev. D45 (1992) 986.
   \bibitem{PDB}   K.A. Olive et al. (Particle Data Group), Chin. Phys. C, 38, 090001 (2014).
  \bibitem{Rpi-scalar}  G. Colangelo, J. Gasser, H. Leutwyler, Nucl.Phys.B603,125(2001).
  \bibitem{Lattice}Vera Gulpers, Georg von Hippel, Hartmut Wittig,  
 arXiv:1507.01749 (July 2015)[hep-lat]
 \bibitem{Reta-A2} P. Aguar-Bartolome  et al. (Mainz A2 Collaboration) Phys.Rev. C89, 044608(2014).
 \bibitem{Reta-disp} C. Hanhart et al., Eur. J. C73, 2668(2013).
 \bibitem{Kubis} Bastian Kubis and Judith Plenter, Eur. J. C75, 283(2015). 
 \bibitem{R-pS} J. Gasser, H. Leutwyler, M.E. Sainio, Phys.Lett. B253,260 (1991).
 \bibitem{A2} M.Ostrick, private communication
 \bibitem{Mike} M.Williams, private communication. 
 \bibitem{VMD} C.Crawford et al.,Phys.Rev. C82, 045211(2010).Earle L. Lomon, Simone Pacetti, Phys.Rev. D85 (2012) 113004, Phys.Rev. D86 (2012) 039901
 \bibitem{anomaly} J.S. Bell and R. Jaciw, Nuovo Cimento 60A, 47 
(1969). S.L. Adler, Phys. Rev. 177, 2426 (1969).
\bibitem{AB-review} A.M.Bernstein and Barry R. Holstein, Rev. Mod. Phys. 85, 49(2013). 
\bibitem{Rory-review} R. Miskimen, Ann. Rev. Nucl. Part. Science, 61, 1 (2011). 
\bibitem{AB} J.Friedrich, private communication.  
 \bibitem{Frascati} D. Babusci et al., Eur. Phys.J. C72,1917 (2012),  Int. J. Mod. Phys. Conf. Ser. 35, 1460395 (2014). 
  \bibitem{Heiri} H. Leutwyler, private communication. 

  \end{thebibliography}
\end{document}